\documentclass[aps,prl,amsmath,amssymb,twocolumn,superscriptaddress,letterpaper]{revtex4}
\usepackage{dcolumn}
\usepackage{graphicx}
\usepackage{epstopdf}
\usepackage{verbatim}
\usepackage{amssymb}
\usepackage{amsmath}
\usepackage{subfigure,wrapfig}
\usepackage{color}

\begin{document}

\title{Structural Colors from Fano Resonances}

\author{Yichen Shen,$^{1,2\star\dagger}$ Veronika Rinnerbauer,$^{2,3\star}$ Imbert Wang,$^{2}$ \\ Veronika Stelmakh,$^{2}$ John D Joannopoulos,$^{1,2}$ and Marin Solja\v{c}i\'{c}$^{1,2}$}
\affiliation{
\normalsize{$^{1}$Department of Physics, Massachusetts Institute of Technology,}\\
\normalsize{Cambridge, MA 02139, USA}\\
\normalsize{$^{2}$Institute For Soldier Nanotechnologies, Massachusetts Institute of Technology,}\\
\normalsize{Cambridge, MA 02139, USA}\\
\normalsize{$^{3}$Institute of Semiconductor and Solid State Physics, Johannes Kepler University,}\\
\normalsize{Linz, Austria}\\
\normalsize{$^\star$ These authors contributed equally to the work}\\
\normalsize{$^\dagger$To whom correspondence should be addressed; E-mail:  ycshen@mit.edu.}
}

\begin{abstract}

Structural coloration is an interference phenomenon where colors emerge when visible light interacts with nanoscopically structured material, and has recently become a most interesting scientific and engineering topic. However, current structural color generation mechanisms either require thick (compared to the wavelength) structures or lack dynamic tunability. This report proposes a new structural color generation mechanism, that produces colors by the Fano resonance effect on thin photonic crystal slab. We experimentally realize the proposed idea by fabricating the samples that show resonance-induced colors with weak dependence on the viewing angle. Finally, we show that the resonance-induced colors can be dynamically tuned by stretching the photonic crystal slab fabricated on an elastic substrate. 

\end{abstract}

\maketitle
People have long been inspired by nature's ability to create a dazzling range of colors. Dyes, pigments and metals that absorb light of only certain colors and reflect the rest of the spectrum are the most common color generation mechanisms. Recently, great attention has been paid to a different type of color generation mechanism: structural coloration, which produces color without the use of dyes and pigments \cite{kinoshita2008physics}. Structural coloration is, in principle, an interference phenomenon where colors emerge when visible light interacts with nanoscopically structured material. Compared with pigmentary colors, structural colors 1) usually appear brighter under sunlight; 2) are immune to photobleaching, and 3) can be tuned dynamically \cite{graham2009tunable}. Due to those advantages, they have found applications in painting, textiles and passive displays \cite{kim2009structural,arsenault2007photonic}. 

Structural colors are typically produced by light interference within bulk materials, for example, structural colors from scattering \cite{prum1998coherent,forster2010biomimetic,noh2010noniridescent}, multilayer interference \cite{kolle2010stretch,kolle2010mimic,kolle2013bio} and photonic crystals\cite{joannopoulos2011photonic,arsenault2007photonic,michaelis2013generating,aschwanden2006polymeric}. In these bulk effects, the color gets brighter and sharper as one increases the bulk thickness, i.e., number of particles (for scattering), the number of layers (for multilayer interference), or the number of periods (for photonic crystals). As a result, designs based on these mechanisms require the structured materials to be much thicker than the wavelength. However, in certain applications, such as optical coating\cite{macleod2001thin,yeh2005optical}, structures with low thickness are preferred \cite{kats2013nanometre}. Conventional optical coatings use thin film interference, which typically requires metallic substrates in order to achieve enough reflection. Therefore, the structural colors generated through thin film interference are material limited, angular sensitive and cannot be tuned dynamically.

In this report, we propose a new mechanism to generate structural color --- different from any of the current mechanisms mentioned above --- that produces color by interference between the directly reflected light and the guided resonances mode on a \textit{surface structure}. The new mechanism is flexible in material choice, suitable for large area fabrication, and has weak angular dependence. Moreover, it supports dynamic tuning of the reflectance spectrum, as we will demonstrate in this study.

\begin{figure*}[htbp]
\begin{center}
\includegraphics[width=6in]{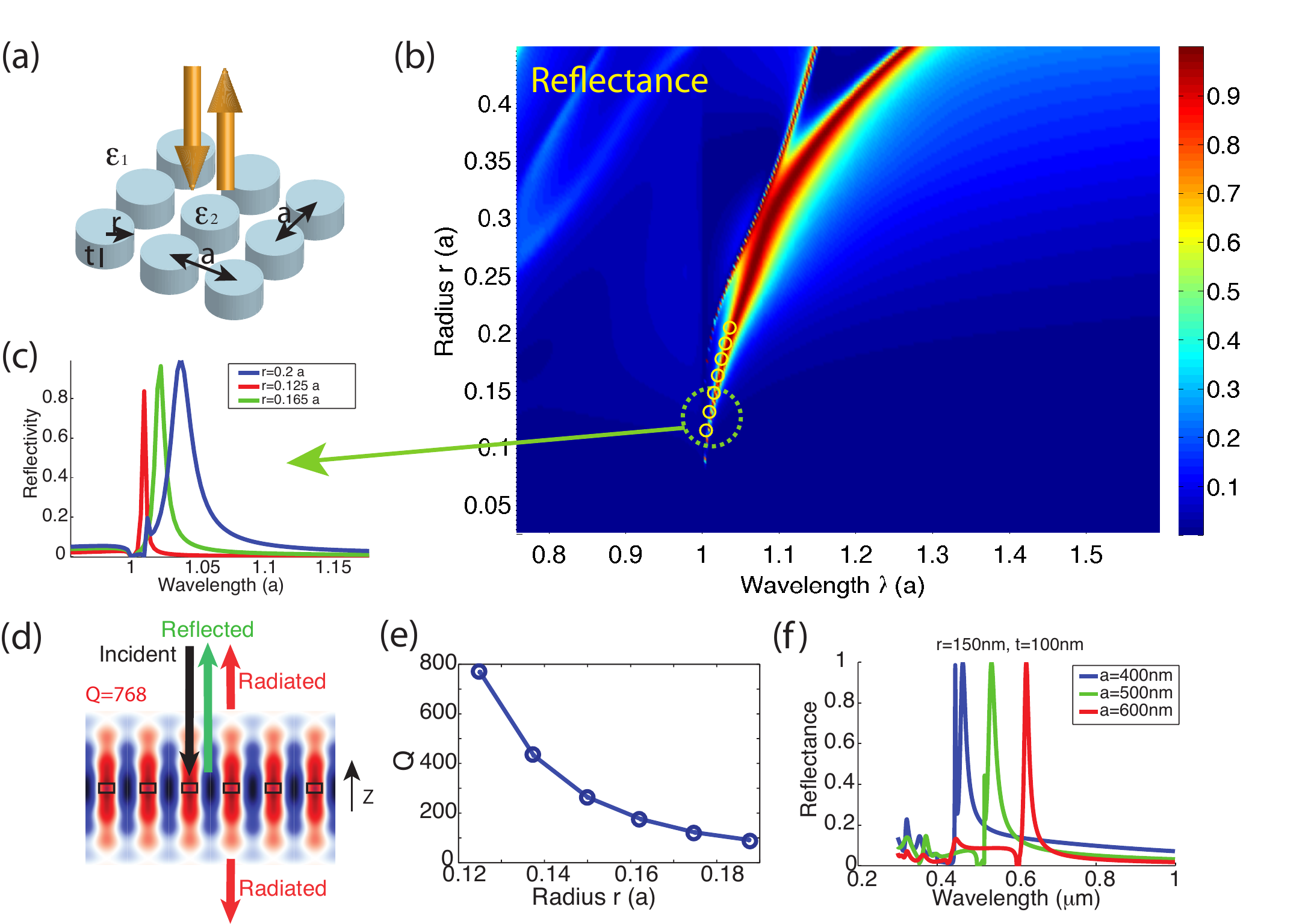}
\caption{\textbf{Light reflectance assisted with Fano-resonance} (a) Light with random polarization incident on a periodic array of nano-rods. The dielectric constant of the nano-rods is set to be $\epsilon_2=4$ (Ta$_2$O$_5$) and the surrounding medium $\epsilon_1=1$ (vacuum). (b) Reflectance spectra versus different radii $r$ of the rods with height $t=0.25a$. The reflectance peaks are the result of interference between the directly reflected light and radiated light from the resonant modes. The yellow rings correspond to the locations of the theoretically simulated resonant modes. (c) Example of reflectance peaks caused by Fano-resonance for structures with $r=0.125a$, $r=0.165a$ and $r=0.2a$. (d) Vertical slice of the spatial distribution of the electric field for the resonant mode responsible for the reflectance peak with $r=0.125a$ in (c). (e) Q values for the resonance modes versus the rod radius. (f) Shift of the reflectance peak when periodicity $a$ is varied, while the rod radius $r=150$nm and rod height $t=100$nm are fixed.}
\label{fig:fig1}
\end{center}
\end{figure*}

The resonance-induced reflectance used in this study to generate static and dynamic color is related to a more general resonance category, known as Fano resonance \cite{miroshnichenko2010fano}. This optical resonance was first investigated and analyzed by Fan \textit{et al} \cite{fan2002analysis} in photonic crystal slabs. The optical Fano resonance can be understood as an interference effect: light incident on a periodic surface structure (Fig.~1a) excites a localized mode supported by the surface structure. The localized mode leaks into the surrounding environment (Fig.~1d), interfering with the directly reflected light from the surface. When the reflected and the radiated light have the same phase, constructive interference produces a sharp reflectance peak (Fig.~1b,c). The reflectance line shape near resonance peak can be calculated by temporal coupled mode theory and fitted by the Lorentzian line shape \cite{fan2003temporal}:
\begin{equation}
r=r_d+f\frac{\gamma}{i(\omega-\omega_0)+\gamma},
\end{equation}
where the factor $f$ is the complex amplitude of the resonant mode, $r_d$ is the direct transmission coefficients, $\omega_0$ and $\gamma$ are the center frequencies and widths of the resonance, which are directly related to the quality factor $Q$ of the resonance mode:
\begin{equation}
Q\approx\frac{\omega_0}{\gamma}.
\end{equation}

To date, most Fano resonance features discussed have a rather high $Q$ ($>50$) to be used for lasing or filtering purposes. In these cases, the reflectance peak is relatively narrow band. However, in order to achieve color control using Fano resonance, the reflectance peak needs to be reasonably broadband to reflect enough light. As a proof of principle, we study a square lattice of rods as shown in Fig.~1a. This structure is also chosen for its suitability for dynamic tuning, which will be discussed later in this paper. The dielectric constant of the rods is chosen to be $\epsilon_2=4$ (e.g., Ta$_2$O$_5$) and the surrounding environment has $\epsilon_1=1$ i.e., vacuum. The reflectance spectra for structures with different rod radii $r$ are computed with the rigorous coupled wave analysis (RCWA) method \cite{Liu20122233} and plotted in Fig.~1b. The bandwidth of the resonant peaks increases with increasing radius $r$ of the rods. The resonant modes at different $r$ and the corresponding field distribution are calculated using a Finite Difference Time Domain (FDTD) method \cite{kunz1993finite} (Fig.~1d) and their locations are highlighted in Fig.~1b. The spectral position of the resonant modes exactly follows the reflectance maxima, suggesting that the reflectance peaks are indeed caused by Fano resonance. The quality factor $Q$ for each resonant mode is evaluated using a low-storage ``filter diagonalization method" (FDM) \cite{mandelshtam1997harmonic} and plotted in Fig.~1e, which shows that $Q$ decreases as $r$ increases.

Furthermore, since the resonant mode of the photonic crystal is directly related to the periodicity of the lattice, one can actively tune the location of the reflectance peak simply by changing the period $a$ of the structure (Fig.~1f).

As a proof of concept, we first realize samples that have resonance-induced, \textit{static} color in the red-to-NIR regime. In the next step, we experimentally realize the same structures on elastic substrates (PDMS) and demonstrate their \textit{dynamic} tunability by varying the periodicity $a$.

The target structure to achieve Fano resonance consists of rods of a high-index material (Si) on a transparent substrate. In our first experiment, the structures proposed in Fig.~1 were fabricated on amorphous SiO$_2$ (glass) substrates, with two different periods of 450~nm and 500~nm, radius from $r=115$ to $r=125$ nm, and height $t=65$ nm. The radius and the thickness of the rods were optimized to support a resonant mode with a $Q$ low enough to produce a significant color effect. In order to get a good optical effect, it is critical that the rods have a high index of refraction while having low absorptivity. That is why amorphous Si was used for the rods.

\begin{figure}[htbp]
\begin{center}
\includegraphics[width=3.4in]{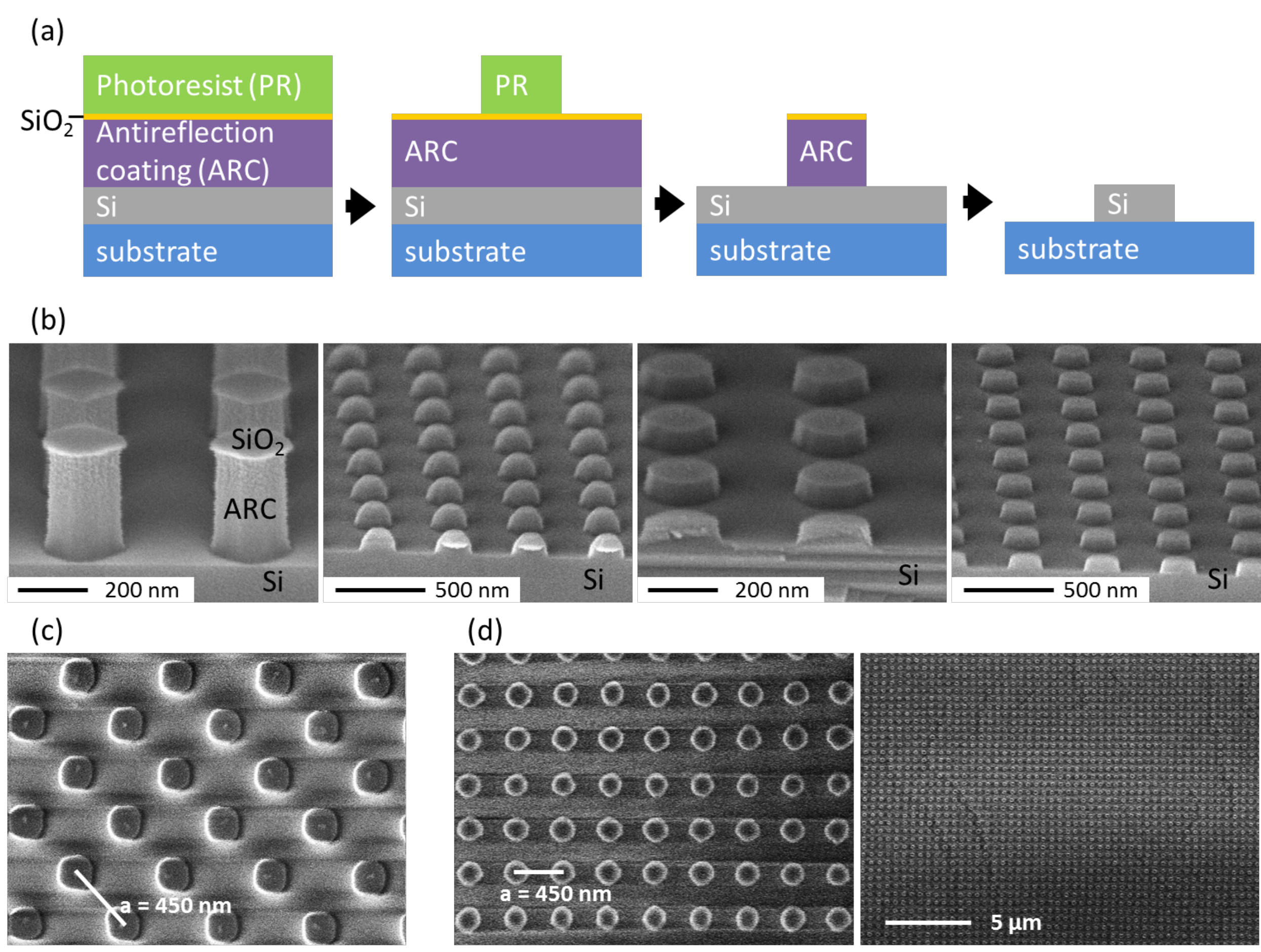}
\caption{\textbf{Fabrication process:}(a) Schematic outline of the process flow: Deposition of the Si layer and lithography layers, pattern definition by interference lithography and pattern transfer by RIE. (b) Scanning electron micrographs of the process steps: Sample after RIE of the ARC, after RIE of the Si layer and after ashing of the remaining ARC (from left to right), cross sections taken from test samples on Si substrates. (c) Final Si rods on glass substrates and on (d) PDMS.}
\label{fig:fig2}
\end{center}
\end{figure}

For the static color samples, fused silica substrates were cleaned and a layer of 65~nm of Si was deposited by plasma enhanced chemical vapor deposition (PECVD) at a substrate temperature of 200$^{\circ}$C. The refractive index and thickness of the Si layer were confirmed by spectroscopic ellipsometry to be $n=4.3$ at 600~nm and 4.0 at 700~nm, while the imaginary part is below $\kappa=0.4$ at 600~nm and vanishes at 700~nm.

The fabrication process of the periodic structures was based on interference lithography (IL) using a trilayer process \cite{schattenburg1995optically,savas1996large} as shown in Fig.~2. IL is a relatively inexpensive, fast and scalable maskless lithography method, which relies on the interference pattern generated by two coherent light sources to define 1D and 2D periodic patterns in a single plane, and is easily scalable to large exposure areas. The lithography layer stack consisted of an antireflective coating (AZ BARLi, 200~nm), a thin layer of SiO$_2$ (20~nm, deposited by electron beam evaporation) to protect the ARC during reactive ion etching (RIE), and PFi-88 (Sumika, 200~nm) as a negative photoresist. The IL was done using a Lloyd's mirror setup with a 325 nm HeCd Laser. The periodicity of the pattern is defined by the interference angle, and the exposure was performed twice, with the substrates rotated by 90$^{\circ}$ between exposures to create a square array of cylindrical rods in the negative resist. After development of the photoresist and RIE of SiO$_2$ and ARC, the pattern was transferred to the Si layer by RIE using a CF$_4$/O$_2$ chemistry. In the end, the remaining ARC was removed by plasma (O$_2$) ashing.

The fabricated sample appeared red in color over a wide range of viewing angles (Fig.~3(c)). The reflectance of the samples at different incident angles was measured using an ultraviolet-visible spectrophotometer (Cary 500i). Notice the reflection peak at normal incidence splits into two reflection bands at 650~nm and 700~nm, respectively. The upper reflection band corresponds to the Fano-resonance from $p$-polarized modes, while the lower reflection band corresponds to the Fano-resonance from $s$-polarized modes. The reflectance spectrum showed a resonance maximum in the range of approximately 650 - 700~nm over a wide range of incidence angles (Fig.~3(b)). Such near-omnidirectional reflective color would be especially useful for displays \cite{Hsu2014Transparent}. The reflectance spectra at normal incidence of samples fabricated with different periodicity $a$ and rod radius $r$ are shown in Fig.~3(d). The resonance wavelength clearly increases as $r$ and $a$ increases as predicted. Compared with the RCWA simulation result (dashed line), the experimental Fano reflectance peaks are broadened due to the non-radiative loss in the structure, possibly due to material absorption and fabrication inaccuracies, in particular local variations in the geometry. Nevertheless, we observed a 3-fold increase in sample's reflectance around the resonance wavelength. Due to the dramatic increase of the absorption in amorphous Si at smaller wavelengths, this particular structure's Fano resonance effect can only be observed at wavelengths longer than 600~nm. However, with suitable lossless materials one should be able to demonstrate this effect also in the lower wavelength regime.

\begin{figure}[htbp]
\begin{center}
\includegraphics[width=3.4in]{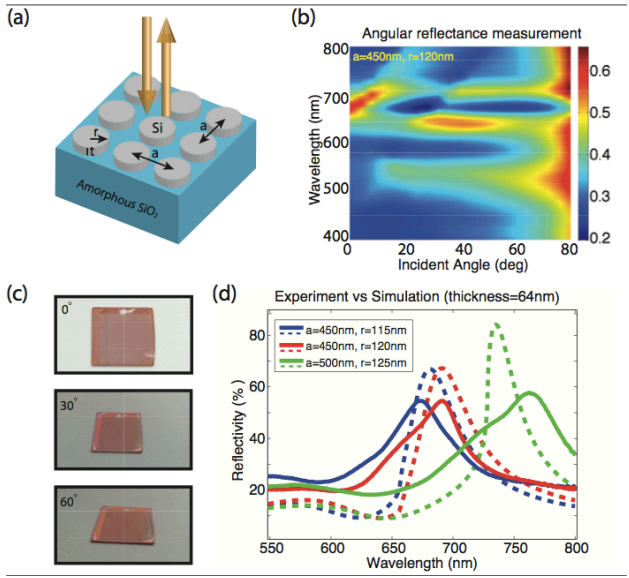}
\caption{\textbf{Si rods on amorphous silica wafer} (a) Illustration of the simulated and fabricated structure. (b) Measured of reflectance spectra of a sample with $a=450$nm and $r=120$nm over all viewing angles. (c) Photographs of the fabricated sample with $a=450$nm and $r=120$nm at different viewing angles. The sample's color appeared red irrespective of the viewing angles (d) Experimental demonstration of color control: Measured (solid lines) and simulated (dashed lines) reflectance spectra of structures with different geometries of the periodic Si rods fabricated on amorphous silica substrates. }
\label{fig:fig3}
\end{center}
\end{figure}

In our second experiment, we demonstrated dynamic color tuning using the same Fano resonance mechanism. In order to achieve the desired effect, the structures were fabricated on polydimethylsiloxane (PDMS) substrates (Fig.~4) instead of glass substrates. To this end, PDMS (Sigma-Aldrich, approx. 200 $\mu$m thick) was spin-coated on 5~cm silanized Si wafers, and annealed at 150$^{\circ}$C for 30 minutes. The Si substrates serve as a support in nanofabrication, and can be removed afterwards by peeling off the thick PDMS layer. The PDMS was surface treated by O$_2$ ashing (200 W) to enhance adhesion and resistance to wet chemical processing, which resulted in some surface damage and macroscopic cracking in the PDMS surface. On the prepared PDMS substrates, a layer of 65~nm of Si was deposited by PEVCD at a substrate temperature of 150$^{\circ}$C.

The fabrication process of the Si rods on the prepared PDMS substrates was similar to that on the glass substrates (Section 3). To achieve higher fabrication accuracy, a Mach Zehnder setup was used for the IL on the PDMS substrates, instead of the Lloyd's mirror.\cite{walsh2004design,anderson1988fabrication,schattenburg1990x}.

In the fabrication process it must be taken into account that PDMS is temperature sensitive and elastic, and also highly sensitive to wet chemical processing. While the deposition of Si on glass was straightforward, the PECVD on PDMS resulted in a dull Si surface, on which the cracks from ashing were plainly visible, with considerable surface roughness visible under a microscope. Nevertheless, the Si layer on PDMS appears to be a sufficiently flat layer on a nanoscopic scale, and the optical properties are close to those of amorphous Si. As a result of the difficult fabrication process on the elastic substrates, the achieved structures have an acceptable nanoscopic accuracy (Fig.~2), but rather low macroscopic uniformity.

In order to achieve isotropic variation of the periodicity $a$, the PDMS sample must be stretched isotropically. To this end, the sample was fixed on a black balloon with radius $R_1$ (Fig.~4(b)) much larger than the sample size. The balloon was then gradually expanded to $R_2$. Through this method, the periodicity of the sample was isotropically increased from $a_1$ to $a_2=a_1\cdot\frac{R_2}{R_1}$. The reflectance spectrum at normal incidence was measured with the same ultraviolet-visible photospectrometer as the one used to measure the static colors, and is shown in Fig.~4(d). Compared with the RCWA simulation result (dashed line in Fig.~4(c)), which was obtained using the measured refractive index dispersion of the deposited Si and assuming a constant rod radius, the spectral position of the experimental reflectance peaks and the simulated spectra match well. The spectral shift of the measured reflectance peak was 32 nm when the sample is stretched by approx.\ 10 \%. The magnitude of the measured reflectance is lower than simulated due to the fabrication inaccuracies mentioned above. Since the Si layer on PDMS was not quite flat after Si deposition, interference lithography was problematic: there was some variation in rod size over a length scale of a couple of periods, together with a variation of the level of the rods, which could have had an adverse effect on the observation of Fano resonance. Also, the RIE of the Si layer and the mask layer removal by ashing might have damaged the PDMS surface to some extent and led to the formation of SiO$_2$ on the PDMS surface. 

\begin{figure}[htbp]
\begin{center}
\includegraphics[width=3.4in]{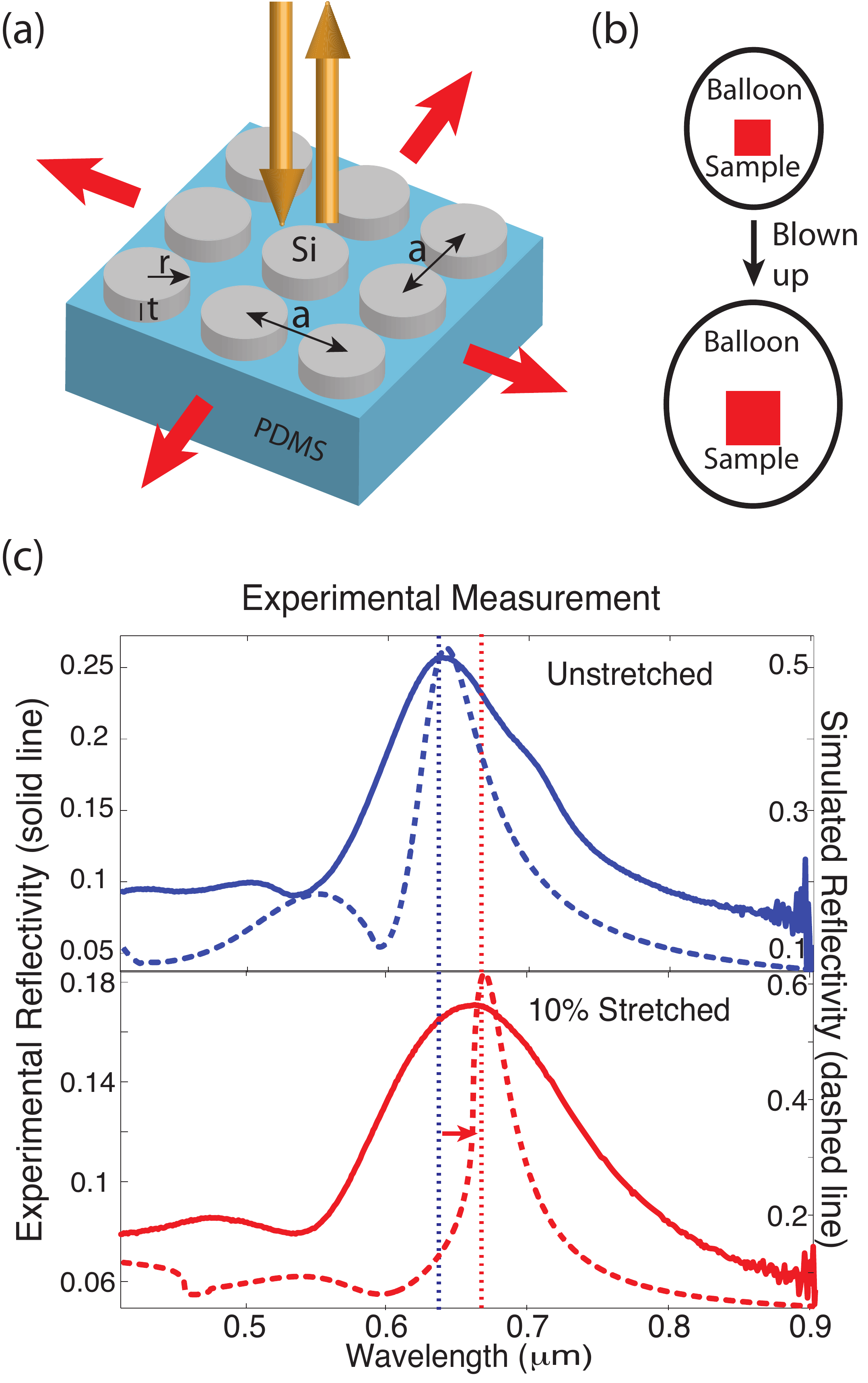}
\caption{\textbf{Si rods on PDMS} (a) Illustration of the simulated and fabricated structure. (b) The periodicity $a$ of the rod lattice can be adjusted by stretching the PDMS. In order to simulate isotropic stretching, the sample is fixed on a balloon, which is gradually blown up. (c) Dynamic color control: Measured (solid lines) and simulated (with ideal structure, dashed lines) reflectance spectra of silicon rods with $a=450$nm and $r=100$nm on polydimethylsiloxane (PDMS) before and after stretching by 10\%.}
\label{fig:fig4}
\end{center}
\end{figure}

In this letter, we propose a structural color mechanism that uses the optical Fano resonance effect of light on a periodic surface structure. To this end, we have designed and fabricated a structure of periodic high-index rods on low-index substrates that show Fano resonance of the reflected and guided modes in the visible wavelength range. The predicted Fano resonance maxima were observed in the reflectance spectra on glass and PDMS substrates. Moreover, dynamical color control was achieved by isotropically stretching the elastic PDMS substrates with the fabricated Fano structure, resulting in a shift of reflection peaks in agreement with with simulation. 

This mechanism for structural color and the geometry demonstrated is very versatile and can be specifically tailored for different applications e.g.\ in displays or light emitting devices. In particular, the Q-factor and therefore the bandwidth of the resonance can be tuned to achieve a broad reflection bandwidth for displays or narrow reflectance bandwidth for light emitting or lasing applications. In addition, the designed structure shows good performance over a broad range of viewing angles. 

The proposed method has the potential to lead the way to static and dynamic color control for a broad range of applications.
The resonance effect achieved with these structures can be further optimized by improving the fabrication mechanism to achieve better accuracy and uniformity. Furthermore, the spectral range over which the effect and the reflectance color can be tuned could be extended by using materials that are lossless in the full visible spectrum, such as TiO$_2$ or Ta$_2$O$_5$.

We thank C. W. Hsu, D. Jukic for valuable discussion. This work was partially supported by the Army Research Office through the ISN under Contract Nos.~W911NF-13-D0001. M.S. were (partially) supported by the MIT S3TEC Energy Research Frontier Center of the Department of Energy under Grant No. DE-SC0001299. V.R. gratefully acknowledges support by the Austrian Science Fund (FWF) J3161-N20.

\bibliography{Yichen_bib}

\begin{thebibliography}{28}
\expandafter\ifx\csname natexlab\endcsname\relax\def\natexlab#1{#1}\fi
\expandafter\ifx\csname bibnamefont\endcsname\relax
  \def\bibnamefont#1{#1}\fi
\expandafter\ifx\csname bibfnamefont\endcsname\relax
  \def\bibfnamefont#1{#1}\fi
\expandafter\ifx\csname citenamefont\endcsname\relax
  \def\citenamefont#1{#1}\fi
\expandafter\ifx\csname url\endcsname\relax
  \def\url#1{\texttt{#1}}\fi
\expandafter\ifx\csname urlprefix\endcsname\relax\def\urlprefix{URL }\fi
\providecommand{\bibinfo}[2]{#2}
\providecommand{\eprint}[2][]{\url{#2}}

\bibitem[{\citenamefont{Kinoshita et~al.}(2008)\citenamefont{Kinoshita,
  Yoshioka, and Miyazaki}}]{kinoshita2008physics}
\bibinfo{author}{\bibfnamefont{S.}~\bibnamefont{Kinoshita}},
  \bibinfo{author}{\bibfnamefont{S.}~\bibnamefont{Yoshioka}}, \bibnamefont{and}
  \bibinfo{author}{\bibfnamefont{J.}~\bibnamefont{Miyazaki}},
  \bibinfo{journal}{Rep. Prog. Phys.} \textbf{\bibinfo{volume}{71}},
  \bibinfo{pages}{076401} (\bibinfo{year}{2008}).

\bibitem[{\citenamefont{Graham-Rowe}(2009)}]{graham2009tunable}
\bibinfo{author}{\bibfnamefont{D.}~\bibnamefont{Graham-Rowe}},
  \bibinfo{journal}{Nat. Photonics} \textbf{\bibinfo{volume}{3}},
  \bibinfo{pages}{551} (\bibinfo{year}{2009}).

\bibitem[{\citenamefont{Kim et~al.}(2009)\citenamefont{Kim, Ge, Kim, Choi, Lee,
  Lee, Park, Yin, and Kwon}}]{kim2009structural}
\bibinfo{author}{\bibfnamefont{H.}~\bibnamefont{Kim}},
  \bibinfo{author}{\bibfnamefont{J.}~\bibnamefont{Ge}},
  \bibinfo{author}{\bibfnamefont{J.}~\bibnamefont{Kim}},
  \bibinfo{author}{\bibfnamefont{S.-e.} \bibnamefont{Choi}},
  \bibinfo{author}{\bibfnamefont{H.}~\bibnamefont{Lee}},
  \bibinfo{author}{\bibfnamefont{H.}~\bibnamefont{Lee}},
  \bibinfo{author}{\bibfnamefont{W.}~\bibnamefont{Park}},
  \bibinfo{author}{\bibfnamefont{Y.}~\bibnamefont{Yin}}, \bibnamefont{and}
  \bibinfo{author}{\bibfnamefont{S.}~\bibnamefont{Kwon}},
  \bibinfo{journal}{Nat. Photonics} \textbf{\bibinfo{volume}{3}},
  \bibinfo{pages}{534} (\bibinfo{year}{2009}).

\bibitem[{\citenamefont{Arsenault et~al.}(2007)\citenamefont{Arsenault, Puzzo,
  Manners, and Ozin}}]{arsenault2007photonic}
\bibinfo{author}{\bibfnamefont{A.~C.} \bibnamefont{Arsenault}},
  \bibinfo{author}{\bibfnamefont{D.~P.} \bibnamefont{Puzzo}},
  \bibinfo{author}{\bibfnamefont{I.}~\bibnamefont{Manners}}, \bibnamefont{and}
  \bibinfo{author}{\bibfnamefont{G.~A.} \bibnamefont{Ozin}},
  \bibinfo{journal}{Nat. Photonics} \textbf{\bibinfo{volume}{1}},
  \bibinfo{pages}{468} (\bibinfo{year}{2007}).

\bibitem[{\citenamefont{Prum et~al.}(1998)\citenamefont{Prum, Torres,
  Williamson, and Dyck}}]{prum1998coherent}
\bibinfo{author}{\bibfnamefont{R.~O.} \bibnamefont{Prum}},
  \bibinfo{author}{\bibfnamefont{R.~H.} \bibnamefont{Torres}},
  \bibinfo{author}{\bibfnamefont{S.}~\bibnamefont{Williamson}},
  \bibnamefont{and} \bibinfo{author}{\bibfnamefont{J.}~\bibnamefont{Dyck}},
  \bibinfo{journal}{Nature} \textbf{\bibinfo{volume}{396}}, \bibinfo{pages}{28}
  (\bibinfo{year}{1998}).

\bibitem[{\citenamefont{Forster et~al.}(2010)\citenamefont{Forster, Noh, Liew,
  Saranathan, Schreck, Yang, Park, Prum, Mochrie, O'Hern
  et~al.}}]{forster2010biomimetic}
\bibinfo{author}{\bibfnamefont{J.~D.} \bibnamefont{Forster}},
  \bibinfo{author}{\bibfnamefont{H.}~\bibnamefont{Noh}},
  \bibinfo{author}{\bibfnamefont{S.~F.} \bibnamefont{Liew}},
  \bibinfo{author}{\bibfnamefont{V.}~\bibnamefont{Saranathan}},
  \bibinfo{author}{\bibfnamefont{C.~F.} \bibnamefont{Schreck}},
  \bibinfo{author}{\bibfnamefont{L.}~\bibnamefont{Yang}},
  \bibinfo{author}{\bibfnamefont{J.-G.} \bibnamefont{Park}},
  \bibinfo{author}{\bibfnamefont{R.~O.} \bibnamefont{Prum}},
  \bibinfo{author}{\bibfnamefont{S.~G.} \bibnamefont{Mochrie}},
  \bibinfo{author}{\bibfnamefont{C.~S.} \bibnamefont{O'Hern}},
  \bibnamefont{et~al.}, \bibinfo{journal}{Adv. Mater.}
  \textbf{\bibinfo{volume}{22}}, \bibinfo{pages}{2939} (\bibinfo{year}{2010}).

\bibitem[{\citenamefont{Noh et~al.}(2010)\citenamefont{Noh, Liew, Saranathan,
  Mochrie, Prum, Dufresne, and Cao}}]{noh2010noniridescent}
\bibinfo{author}{\bibfnamefont{H.}~\bibnamefont{Noh}},
  \bibinfo{author}{\bibfnamefont{S.~F.} \bibnamefont{Liew}},
  \bibinfo{author}{\bibfnamefont{V.}~\bibnamefont{Saranathan}},
  \bibinfo{author}{\bibfnamefont{S.~G.} \bibnamefont{Mochrie}},
  \bibinfo{author}{\bibfnamefont{R.~O.} \bibnamefont{Prum}},
  \bibinfo{author}{\bibfnamefont{E.~R.} \bibnamefont{Dufresne}},
  \bibnamefont{and} \bibinfo{author}{\bibfnamefont{H.}~\bibnamefont{Cao}},
  \bibinfo{journal}{Adv. Mater.} \textbf{\bibinfo{volume}{22}},
  \bibinfo{pages}{2871} (\bibinfo{year}{2010}).

\bibitem[{\citenamefont{Kolle et~al.}(2010{\natexlab{a}})\citenamefont{Kolle,
  Zheng, Gibbons, Baumberg, and Steiner}}]{kolle2010stretch}
\bibinfo{author}{\bibfnamefont{M.}~\bibnamefont{Kolle}},
  \bibinfo{author}{\bibfnamefont{B.}~\bibnamefont{Zheng}},
  \bibinfo{author}{\bibfnamefont{N.}~\bibnamefont{Gibbons}},
  \bibinfo{author}{\bibfnamefont{J.~J.} \bibnamefont{Baumberg}},
  \bibnamefont{and} \bibinfo{author}{\bibfnamefont{U.}~\bibnamefont{Steiner}},
  \bibinfo{journal}{Opt. Express} \textbf{\bibinfo{volume}{18}},
  \bibinfo{pages}{4356} (\bibinfo{year}{2010}{\natexlab{a}}).

\bibitem[{\citenamefont{Kolle et~al.}(2010{\natexlab{b}})\citenamefont{Kolle,
  Salgard-Cunha, Scherer, Haung, Vukusic, Mahajan, Baumberg, and
  Steiner}}]{kolle2010mimic}
\bibinfo{author}{\bibfnamefont{M.}~\bibnamefont{Kolle}},
  \bibinfo{author}{\bibfnamefont{P.~M.} \bibnamefont{Salgard-Cunha}},
  \bibinfo{author}{\bibfnamefont{M.~R.} \bibnamefont{Scherer}},
  \bibinfo{author}{\bibfnamefont{F.}~\bibnamefont{Haung}},
  \bibinfo{author}{\bibfnamefont{P.}~\bibnamefont{Vukusic}},
  \bibinfo{author}{\bibfnamefont{S.}~\bibnamefont{Mahajan}},
  \bibinfo{author}{\bibfnamefont{J.~J.} \bibnamefont{Baumberg}},
  \bibnamefont{and} \bibinfo{author}{\bibfnamefont{U.}~\bibnamefont{Steiner}},
  \bibinfo{journal}{Nat. Nanotechnol.} \textbf{\bibinfo{volume}{5}},
  \bibinfo{pages}{511} (\bibinfo{year}{2010}{\natexlab{b}}).

\bibitem[{\citenamefont{Kolle et~al.}(2013)\citenamefont{Kolle, Lethbridge,
  Kreysing, Baumberg, Aizenberg, and Vukusic}}]{kolle2013bio}
\bibinfo{author}{\bibfnamefont{M.}~\bibnamefont{Kolle}},
  \bibinfo{author}{\bibfnamefont{A.}~\bibnamefont{Lethbridge}},
  \bibinfo{author}{\bibfnamefont{M.}~\bibnamefont{Kreysing}},
  \bibinfo{author}{\bibfnamefont{J.~J.} \bibnamefont{Baumberg}},
  \bibinfo{author}{\bibfnamefont{J.}~\bibnamefont{Aizenberg}},
  \bibnamefont{and} \bibinfo{author}{\bibfnamefont{P.}~\bibnamefont{Vukusic}},
  \bibinfo{journal}{Adv. Mater.} \textbf{\bibinfo{volume}{25}},
  \bibinfo{pages}{2239} (\bibinfo{year}{2013}).

\bibitem[{\citenamefont{Joannopoulos et~al.}(2011)\citenamefont{Joannopoulos,
  Johnson, Winn, and Meade}}]{joannopoulos2011photonic}
\bibinfo{author}{\bibfnamefont{J.}~\bibnamefont{Joannopoulos}},
  \bibinfo{author}{\bibfnamefont{S.}~\bibnamefont{Johnson}},
  \bibinfo{author}{\bibfnamefont{J.}~\bibnamefont{Winn}}, \bibnamefont{and}
  \bibinfo{author}{\bibfnamefont{R.}~\bibnamefont{Meade}},
  \emph{\bibinfo{title}{Photonic Crystals: Molding the Flow of Light (Second
  Edition)}} (\bibinfo{publisher}{Princeton University Press},
  \bibinfo{year}{2011}), ISBN \bibinfo{isbn}{9781400828241},
  \urlprefix\url{http://books.google.com/books?id=owhE36qiTP8C}.

\bibitem[{\citenamefont{Michaelis et~al.}(2013)\citenamefont{Michaelis,
  Snoswell, Bell, Spahn, Hellmann, Finlayson, and
  Baumberg}}]{michaelis2013generating}
\bibinfo{author}{\bibfnamefont{B.}~\bibnamefont{Michaelis}},
  \bibinfo{author}{\bibfnamefont{D.~R.} \bibnamefont{Snoswell}},
  \bibinfo{author}{\bibfnamefont{N.~A.} \bibnamefont{Bell}},
  \bibinfo{author}{\bibfnamefont{P.}~\bibnamefont{Spahn}},
  \bibinfo{author}{\bibfnamefont{G.~P.} \bibnamefont{Hellmann}},
  \bibinfo{author}{\bibfnamefont{C.~E.} \bibnamefont{Finlayson}},
  \bibnamefont{and} \bibinfo{author}{\bibfnamefont{J.~J.}
  \bibnamefont{Baumberg}}, \bibinfo{journal}{Adv. Eng. Mater.}
  \textbf{\bibinfo{volume}{15}}, \bibinfo{pages}{948} (\bibinfo{year}{2013}).

\bibitem[{\citenamefont{Aschwanden and
  Stemmer}(2006)}]{aschwanden2006polymeric}
\bibinfo{author}{\bibfnamefont{M.}~\bibnamefont{Aschwanden}} \bibnamefont{and}
  \bibinfo{author}{\bibfnamefont{A.}~\bibnamefont{Stemmer}},
  \bibinfo{journal}{Opt. Lett.} \textbf{\bibinfo{volume}{31}},
  \bibinfo{pages}{2610} (\bibinfo{year}{2006}).

\bibitem[{\citenamefont{Macleod}(2001)}]{macleod2001thin}
\bibinfo{author}{\bibfnamefont{H.~A.} \bibnamefont{Macleod}},
  \emph{\bibinfo{title}{Thin-film optical filters}} (\bibinfo{publisher}{CRC
  Press}, \bibinfo{year}{2001}).

\bibitem[{\citenamefont{Yeh}(2005)}]{yeh2005optical}
\bibinfo{author}{\bibfnamefont{P.}~\bibnamefont{Yeh}},
  \emph{\bibinfo{title}{Optical Waves in Layered Media}}, Wiley Series in Pure
  and Applied Optics (\bibinfo{publisher}{Wiley}, \bibinfo{year}{2005}), ISBN
  \bibinfo{isbn}{9780471731924},
  \urlprefix\url{http://books.google.com/books?id=-yZBAQAAIAAJ}.

\bibitem[{\citenamefont{Kats et~al.}(2013)\citenamefont{Kats, Blanchard,
  Genevet, and Capasso}}]{kats2013nanometre}
\bibinfo{author}{\bibfnamefont{M.~A.} \bibnamefont{Kats}},
  \bibinfo{author}{\bibfnamefont{R.}~\bibnamefont{Blanchard}},
  \bibinfo{author}{\bibfnamefont{P.}~\bibnamefont{Genevet}}, \bibnamefont{and}
  \bibinfo{author}{\bibfnamefont{F.}~\bibnamefont{Capasso}},
  \bibinfo{journal}{Nat. Mater.} \textbf{\bibinfo{volume}{12}},
  \bibinfo{pages}{20} (\bibinfo{year}{2013}).

\bibitem[{\citenamefont{Miroshnichenko
  et~al.}(2010)\citenamefont{Miroshnichenko, Flach, and
  Kivshar}}]{miroshnichenko2010fano}
\bibinfo{author}{\bibfnamefont{A.~E.} \bibnamefont{Miroshnichenko}},
  \bibinfo{author}{\bibfnamefont{S.}~\bibnamefont{Flach}}, \bibnamefont{and}
  \bibinfo{author}{\bibfnamefont{Y.~S.} \bibnamefont{Kivshar}},
  \bibinfo{journal}{Rev. Mod. Phys.} \textbf{\bibinfo{volume}{82}},
  \bibinfo{pages}{2257} (\bibinfo{year}{2010}).

\bibitem[{\citenamefont{Fan and Joannopoulos}(2002)}]{fan2002analysis}
\bibinfo{author}{\bibfnamefont{S.}~\bibnamefont{Fan}} \bibnamefont{and}
  \bibinfo{author}{\bibfnamefont{J.}~\bibnamefont{Joannopoulos}},
  \bibinfo{journal}{Phys. Rev. B} \textbf{\bibinfo{volume}{65}},
  \bibinfo{pages}{235112} (\bibinfo{year}{2002}).

\bibitem[{\citenamefont{Fan et~al.}(2003)\citenamefont{Fan, Suh, and
  Joannopoulos}}]{fan2003temporal}
\bibinfo{author}{\bibfnamefont{S.}~\bibnamefont{Fan}},
  \bibinfo{author}{\bibfnamefont{W.}~\bibnamefont{Suh}}, \bibnamefont{and}
  \bibinfo{author}{\bibfnamefont{J.}~\bibnamefont{Joannopoulos}},
  \bibinfo{journal}{JOSA A} \textbf{\bibinfo{volume}{20}}, \bibinfo{pages}{569}
  (\bibinfo{year}{2003}).

\bibitem[{\citenamefont{Liu and Fan}(2012)}]{Liu20122233}
\bibinfo{author}{\bibfnamefont{V.}~\bibnamefont{Liu}} \bibnamefont{and}
  \bibinfo{author}{\bibfnamefont{S.}~\bibnamefont{Fan}},
  \bibinfo{journal}{Comput. Phys. Commun.} \textbf{\bibinfo{volume}{183}},
  \bibinfo{pages}{2233 } (\bibinfo{year}{2012}), ISSN
  \bibinfo{issn}{0010-4655},
  \urlprefix\url{http://www.sciencedirect.com/science/article/pii/S0010465512001658}.

\bibitem[{\citenamefont{Kunz and Luebbers}(1993)}]{kunz1993finite}
\bibinfo{author}{\bibfnamefont{K.~S.} \bibnamefont{Kunz}} \bibnamefont{and}
  \bibinfo{author}{\bibfnamefont{R.~J.} \bibnamefont{Luebbers}},
  \emph{\bibinfo{title}{The finite difference time domain method for
  electromagnetics}} (\bibinfo{publisher}{CRC press}, \bibinfo{year}{1993}).

\bibitem[{\citenamefont{Mandelshtam and
  Taylor}(1997)}]{mandelshtam1997harmonic}
\bibinfo{author}{\bibfnamefont{V.~A.} \bibnamefont{Mandelshtam}}
  \bibnamefont{and} \bibinfo{author}{\bibfnamefont{H.~S.}
  \bibnamefont{Taylor}}, \bibinfo{journal}{J. Chem. Phys}
  \textbf{\bibinfo{volume}{107}}, \bibinfo{pages}{6756} (\bibinfo{year}{1997}).

\bibitem[{\citenamefont{Schattenburg et~al.}(1995)\citenamefont{Schattenburg,
  Aucoin, and Fleming}}]{schattenburg1995optically}
\bibinfo{author}{\bibfnamefont{M.}~\bibnamefont{Schattenburg}},
  \bibinfo{author}{\bibfnamefont{R.}~\bibnamefont{Aucoin}}, \bibnamefont{and}
  \bibinfo{author}{\bibfnamefont{R.}~\bibnamefont{Fleming}},
  \bibinfo{journal}{J. Vac. Sci. Technol. B} \textbf{\bibinfo{volume}{13}},
  \bibinfo{pages}{3007} (\bibinfo{year}{1995}).

\bibitem[{\citenamefont{Savas et~al.}(1996)\citenamefont{Savas, Schattenburg,
  Carter, and Smith}}]{savas1996large}
\bibinfo{author}{\bibfnamefont{T.}~\bibnamefont{Savas}},
  \bibinfo{author}{\bibfnamefont{M.}~\bibnamefont{Schattenburg}},
  \bibinfo{author}{\bibfnamefont{J.}~\bibnamefont{Carter}}, \bibnamefont{and}
  \bibinfo{author}{\bibfnamefont{H.~I.} \bibnamefont{Smith}},
  \bibinfo{journal}{J. Vac. Sci. Technol. B} \textbf{\bibinfo{volume}{14}},
  \bibinfo{pages}{4167} (\bibinfo{year}{1996}).

\bibitem[{\citenamefont{Hsu et~al.}(2014)\citenamefont{Hsu, Zhen, Qiu, Shapira,
  DeLacy, Joannopoulos, and Solja{\v c}i{\'c}}}]{Hsu2014Transparent}
\bibinfo{author}{\bibfnamefont{C.~W.} \bibnamefont{Hsu}},
  \bibinfo{author}{\bibfnamefont{B.}~\bibnamefont{Zhen}},
  \bibinfo{author}{\bibfnamefont{W.}~\bibnamefont{Qiu}},
  \bibinfo{author}{\bibfnamefont{O.}~\bibnamefont{Shapira}},
  \bibinfo{author}{\bibfnamefont{B.~G.} \bibnamefont{DeLacy}},
  \bibinfo{author}{\bibfnamefont{J.~D.} \bibnamefont{Joannopoulos}},
  \bibnamefont{and} \bibinfo{author}{\bibfnamefont{M.}~\bibnamefont{Solja{\v
  c}i{\'c}}}, \bibinfo{journal}{Nat. Commun.} \textbf{\bibinfo{volume}{5}},
  \bibinfo{pages}{3152} (\bibinfo{year}{2014}),
  \urlprefix\url{http://dx.doi.org/10.1038/ncomms4152}.

\bibitem[{\citenamefont{Walsh}(2004)}]{walsh2004design}
\bibinfo{author}{\bibfnamefont{M.~E.} \bibnamefont{Walsh}}, Ph.D. thesis,
  \bibinfo{school}{Massachusetts Institute of Technology}
  (\bibinfo{year}{2004}).

\bibitem[{\citenamefont{Anderson}(1988)}]{anderson1988fabrication}
\bibinfo{author}{\bibfnamefont{E.~H.} \bibnamefont{Anderson}}, Ph.D. thesis,
  \bibinfo{school}{Massachusetts Institute of Technology}
  (\bibinfo{year}{1988}).

\bibitem[{\citenamefont{Schattenburg et~al.}(1990)\citenamefont{Schattenburg,
  Anderson, and Smith}}]{schattenburg1990x}
\bibinfo{author}{\bibfnamefont{M.}~\bibnamefont{Schattenburg}},
  \bibinfo{author}{\bibfnamefont{E.}~\bibnamefont{Anderson}}, \bibnamefont{and}
  \bibinfo{author}{\bibfnamefont{H.~I.} \bibnamefont{Smith}},
  \bibinfo{journal}{Phys. Scr.} \textbf{\bibinfo{volume}{41}},
  \bibinfo{pages}{13} (\bibinfo{year}{1990}).

\end{thebibliography}

\end{document}